\documentclass[
preprint,amsmath,amssymb,apl,pra,prb,superscriptaddress]{revtex4-1}

\usepackage{graphics}
\usepackage{amsmath}
\usepackage{amsfonts}
\usepackage{latexsym}
\usepackage{amsbsy}
\usepackage{epstopdf}

\newcommand{\ket}[1]{\vert #1 \rangle}

\usepackage{color}

\usepackage{ulem} 

\newcommand{\dyadic}[1]{{#1}
\setbox0=\hbox{$\scriptstyle\leftrightarrow$}
   \setbox2=\hbox{$#1$}
   \dimen0=.5\wd0 \advance\dimen0 by-.5\wd2
   \advance\dimen0 by-\wd0
   \kern\dimen0
{^{\hbox{$\scriptstyle\leftrightarrow$}}}}

\begin{document}

%
%

\title{Sub-Wavelength Imaging and Field Mapping\\ via EIT and Autler-Townes Splitting In Rydberg Atoms}
\thanks{This work was partially supported by DARPA's QuASAR program. Publication of the U.S. government, not subject to U.S. copyright.}
\author{Christopher~L.~Holloway}
\email{holloway@boulder.nist.gov}
\author{Joshua A. Gordon}
\affiliation{National Institute of Standards and Technology (NIST), Electromagnetics Division,
U.S. Department of Commerce, Boulder Laboratories,
Boulder,~CO~80305}
\author{Andrew Schwarzkopf}
\author{David A. Anderson}
\author{Stephanie A. Miller}
\author{Nithiwadee Thaicharoen}
\author{Georg Raithel}
\affiliation{Department of Physics, University of Michigan, Ann Arbor, MI 48109}

\date{\today}

\begin{abstract}
We present a technique for measuring radio-frequency (RF) electric field strengths with sub-wavelength resolution. We use Rydberg states of rubidium atoms to probe the RF field.  The RF field causes an energy splitting of the Rydberg states via the Autler-Townes effect, and we detect the splitting via electromagnetically induced transparency (EIT).  We use this technique to measure the electric field distribution inside a glass cylinder with applied RF fields at 17.04~GHz and 104.77~GHz. We achieve a spatial resolution of $\bf{\approx}$100~$\bf{\mu}$m, limited by the widths of the laser beams utilized for the EIT spectroscopy. We numerically simulate the fields in the glass cylinder and find good agreement with the measured fields. Our results suggest that this technique could be applied to image fields on a small spatial scale over a large range of frequencies, up into the sub-THz regime.
\vspace{7mm}
\end{abstract}

%

\maketitle

\section{Introduction}

The typical probe (sensor) to measure an electric (E) field has a size on the order of $\lambda/2$ or $\lambda/4$ (where $\lambda$ is the free-space wavelength of the radiation one intends to measure). One example is a dipole loaded field probe \cite{dipole}. These conventional probes
can only measure an E-field strength averaged over the length of the probe. If one is interested in measuring a field distribution in the neighborhood of a structure with spatial features smaller than $\lambda$, the conventional probe would be problematic. For example, consider metasurface structures \cite{apm}.  These metasurfaces are typically composed of periodic arrays of inclusions or scatterers. These scatterers are typically sub-wavelength in size (on the order of $\lambda/10$ or smaller). Furthermore, these scatterers have even smaller sub-structures (gaps, holes,
apertures) that can be as small as $\sim\lambda/100$. With current methods, it is virtually impossible to measure E-field distributions on these sub-wavelength scales.  The technique we demonstrate to perform high spatial resolution mapping of RF fields could be used in design verification and characterization of the metasurfaces and other sub-wavelength devices and structures.

\begin{figure}[!t]
\centering
\scalebox{.5}{\includegraphics*{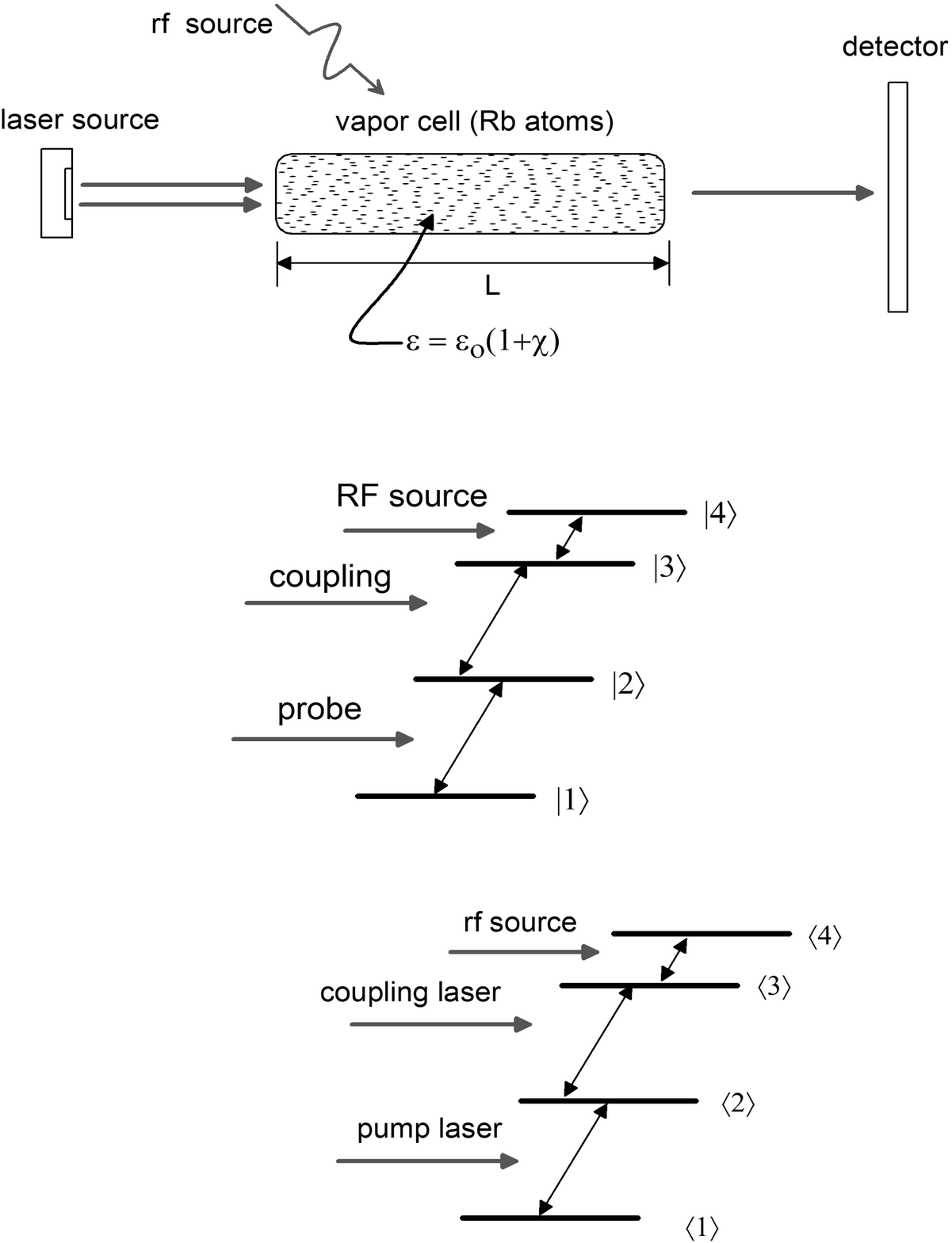}}
\caption{A four-level system.}
\label{4level}
\end{figure}
\begin{figure}
\centering
\scalebox{.45}{\includegraphics*{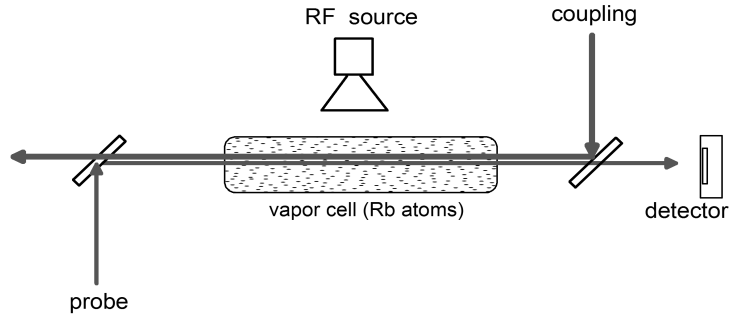}}
\caption{Vapor cell setup for measuring EIT, with counter-propagating probe (red) and coupling (blue) beams. The RF is applied transverse to the optical beam propagation in the vapor cell.}
\label{vaporcell}
\end{figure}

To measure E-fields with sub-wavelength resolution, we take advantage of a recently-demonstrated technique that uses atoms as field probes \cite{jim}-\cite{gordon}.  Here we demonstrate, for the first time, its spatial resolution capability.
The technique uses room-temperature rubidium atoms as probes, exploiting the sensitivity of their high-lying Rydberg states to radio frequency (RF) radiation.  (The term RF is used here to cover the conventional RF, microwave, millimeter wave, and sub-terahertz spectra.) This sensitivity
reflects the large transition matrix elements ($\wp$, on the order of $10^3$ to $10^4 e a_0$) for RF transitions between Rydberg states.
We measure an Autler-Townes splitting \cite{autler} of Rydberg energy levels in these atoms to obtain the RF field
strength.  The energy levels are measured using electromagnetically induced transparency (EIT) \cite{EIT,EIT2}.

Here we describe the physical principles underlying the technique. Consider a sample of stationary four-level atoms illuminated by a single weak (``probe") light field, as depicted in Fig~\ref{4level}.  When the frequency of the light matches the $\ket{1}$ to $\ket{2}$ atomic resonance, the atoms scatter light from the incident beam and reduce the transmitted light intensity.  If a second strong (``coupling") light field is applied resonant with the $\ket{2}$ to $\ket{3}$ transition, the $\ket{2}$ and $\ket{3}$ states are mixed to form dressed states which are close in energy.
The excitation amplitudes from $\ket{1}$ to each of these two dressed states then have opposite signs, leading to destructive quantum interference of these excitation pathways. As such, a transparency window is opened for the probe light: probe light transmission is increased. This is the phenomenon known as EIT \cite{EIT}. If one applies an RF field which couples states $\ket{3}$ and $\ket{4}$, a third dressed state is introduced between the two involved in EIT which leads to constructive interference in the probe absorption.  This splits the EIT resonance in two, and for resonant driving fields the new transmission maxima are split by the Rabi frequency $\Omega_{RF}$ of the $\ket{3}$ -- $\ket{4}$ transition \cite{Lukin1999,Dutta2007}. This is known as Autler-Townes splitting of the EIT signal \cite{tony}.  Therefore the frequency splitting $\Delta f_0$ between the transmission maxima allows a measurement of the E-field amplitude of the RF field via
\begin{equation}
	|E|=\frac{\hbar}{\wp}{\Omega_{RF}}=2\pi\frac{\hbar}{\wp}{\Delta f_0}
	\label{mage}
\end{equation}
where $\hbar$ is Planck's constant and $\wp$ is the electric dipole moment of the RF transition $\ket{3}$ to $\ket{4}$.

In order to measure the field amplitude for different RF frequencies, different states $\ket{3}$ and $\ket{4}$ can be chosen.  State $\ket{3}$ is selected by tuning the wavelength of the coupling laser.  A large range of atomic transitions can be selected, allowing measurements of microwave fields over a correspondingly wide selection of frequencies.
In essence, the atoms act as highly-tunable, resonant RF detectors.  This is a significant benefit of using Rydberg atoms as field probes.
In \cite{broadband}, we use this fact to show how the technique can be used for broadband measurements of RF E-fields, ranging from 1~GHz to 500~GHz.

\section{Experimental Setup}
\begin{figure}[!t]
\centering
\scalebox{.3}
{\includegraphics*{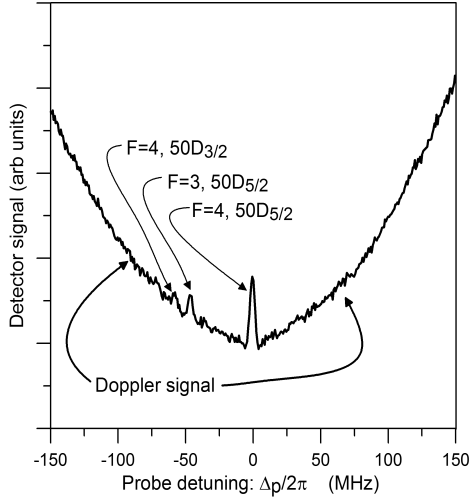}}
\caption{Probe transmission as a function of $\Delta_p$ for the three-level $5S_{1/2}-5P_{3/2}-50D$ EIT system.  EIT peaks are visible for transitions corresponding to (right to left) $5P_{3/2} (F=4)$ -- $50D_{5/2}$, $5P_{3/2} (F=3)$ -- $50D_{5/2}$, and $5P_{3/2} (F=4)$ -- $50D_{3/2}$.}
\label{doppler}
\end{figure}

The experimental setup is shown in Fig.~{\ref{vaporcell}}. We use a cylindrical glass vapor cell of length 75~mm and diameter 25~mm containing rubidium-85 ($^{85}$Rb) atoms.  The levels $\ket{1}$, $\ket{2}$, $\ket{3}$, and $\ket{4}$ correspond respectively to the $^{85}$Rb  $5S_{1/2}$ ground state,  $5P_{3/2}$ excited state, and two Rydberg states.  The probe
is a 780~nm laser (``red") which is scanned across the $5S_{1/2}$ -- $5P_{3/2}$ transition.  The probe beam is focused to a full-width at half maximum (FWHM) of 80~$\mu$m, with a power of 120~nW to keep the intensity below the saturation intensity of the transition.
Figure~\ref{doppler} shows a typical transmission signal as a function of relative probe detuning $\Delta_p$.
The global shape of the curve is the Doppler absorption spectrum of $^{85}$Rb at room temperature.
To produce an EIT signal, we apply a counter-propagating coupling laser (wavelength $\lambda_c \approx 480$~nm, ``blue") with a power of 22~mW, focused to a FWHM of 100~$\mu$m.
As an example, tuning the coupling laser near the $5P_{3/2}$ -- $50D$ Rydberg transition results in distinct EIT transmission peaks corresponding to the transitions from $5S_{1/2}$ to the allowed $5P_{3/2}$ hyperfine sublevels ($F=4,3,2$), which are strongly coupled to the fine-structure-split $50D_{5/2}$ and $50D_{3/2}$ Rydberg states. The peaks for the strongest of these cascades are visible atop the Doppler profile in Fig~\ref{doppler}.
Differential Doppler shifts between the probe and coupling beams alter the frequency separations between EIT peaks in the probe transmission spectrum.
Splittings of $5P_{3/2}$ hyperfine states are scaled by $1-\lambda_c/\lambda_p$, while splittings of Rydberg states are scaled by $\lambda_c/\lambda_p$ \cite{EIT_Adams}.  The latter factor is relevant to measurements of RF-induced splittings of EIT peaks.

\begin{figure}
\centering
\scalebox{.3}
{\includegraphics*{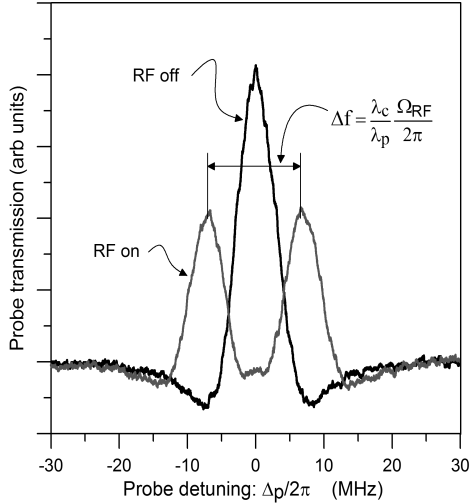}}
\caption{Black curve: EIT-signal as a function of $\Delta_p$ for the EIT system $5S_{1/2} - 5P_{3/2} - 28D_{5/2}$. Gray curve: The $28D_{5/2}$ level is coupled to the $29P_{3/2}$ level by a 104.77~GHz RF field.}
\label{eitsignal}
\end{figure}

Hereafter, for each EIT system we investigate, we focus on the strongest peak, corresponding to the $5P_{3/2} (F=4)$ -- $nD_{5/2}$ transition.  We take this peak to be at $\Delta_p=0$.
In order to improve the signal-to-noise ratio, we use heterodyne detection.  We modulate the blue laser amplitude with a 30~kHz square wave and detect any resulting modulation of the probe transmission with a lock-in amplifier.
This removes the Doppler background and isolates the EIT signal as shown in the black curve of Fig.~\ref{eitsignal}.
Here we tune the coupling laser near the $5P_{3/2}$ -- $28D_{5/2}$ transition.  Application of RF to couple states $28D_{5/2}$ and $29P_{3/2}$ splits the EIT peak as shown in the gray curve.  We measure the frequency splitting of the EIT peaks in the probe spectrum, $\Delta f$, and determine the E-field amplitude by
\begin{equation}
	|E| = 2 \pi \frac{\hbar}{\wp} \frac{\lambda_p}{\lambda_c} \Delta f \quad.
	\label{mage2}
\end{equation}
Note here the use of the Doppler scaling factor, not present in Eq.~\ref{mage} for stationary atoms.

The E-field sensing volume is determined by the overlap of the RF, probe beam, and coupling beam within the vapor cell.
Based on the geometries given above, this volume is a cylinder of length 75~mm and diameter 80~$\mu$m.
The small optical beam diameter gives the measurement high spatial resolution in the dimensions transverse to the optical beams, which is crucial in the experiments presented next.


\section{Sub-Wavelength Field Mapping}

\begin{figure}[!t]
\centering
\scalebox{.5}
{\includegraphics*{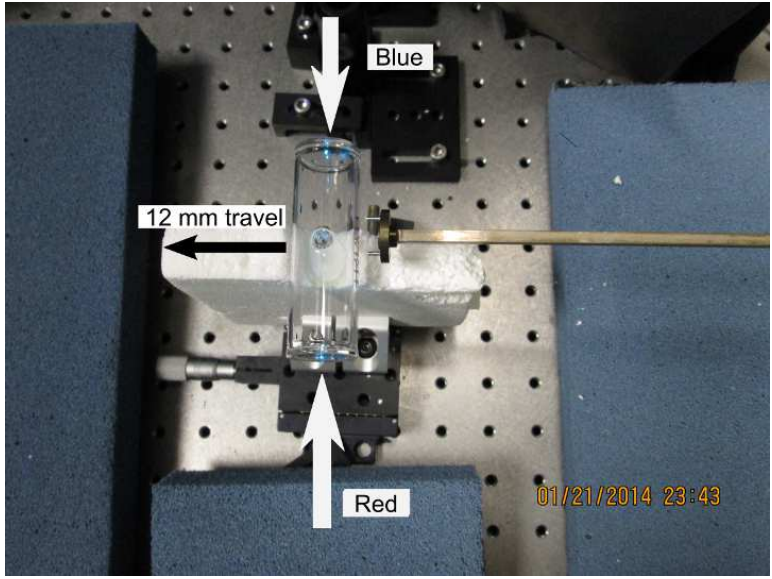}}
\caption{Experimental setup for field-mapping measurements with EIT. The vapor cell is on a translation stage and is scanned with respect to the optical beams. The waveguide in the photo is closer than that used in the measurement.}
\label{exper}
\end{figure}

When an electromagnetic (EM) wave is incident onto a hollow dielectric cylinder, standing waves typically develop inside the cylinder due to internal reflections from the cylinder walls. The resulting field distribution will vary depending on the EM frequency. Using the technique explained in the previous section, we image the field inside our glass vapor cell (which is, electromagnetically, a hollow dielectric cylinder) for two different RF frequencies: 104.77~GHz and 17.04~GHz.  For the 104.77~GHz measurements, we deliver the RF with an open-ended waveguide (see Fig.~\ref{exper}); for 17.04~GHz we use a horn antenna. In each of these measurements, the vapor cell is placed on a translation stage with 12~mm of travel, see Fig.~\ref{exper}. The cell is then translated in a direction perpendicular to the propagation directions of the optical beams.  This allows the imaging of RF fields inside the cell as a function of the spatial coordinate parallel to the translation axis.  The spatial resolution is limited by the optical beam diameter (80~$\mu$m).

We first perform measurements at 104.77~GHz.
The blue laser is tuned to $\approx 482.23$~nm to couple states $5{\rm P}_{3/2}$ and $28{\rm D}_{5/2}$, and the 104.77~GHz field couples $28{\rm D}_{5/2}-29{\rm P}_{3/2}$.
The open-ended waveguide is spaced 0.14~m from the focal axis of the lasers, and is supplied with 0.58~mW of RF power (measured with a power meter attached to the end of the waveguide).

\begin{figure}[!t]
\centering
\scalebox{.4}
{\includegraphics*{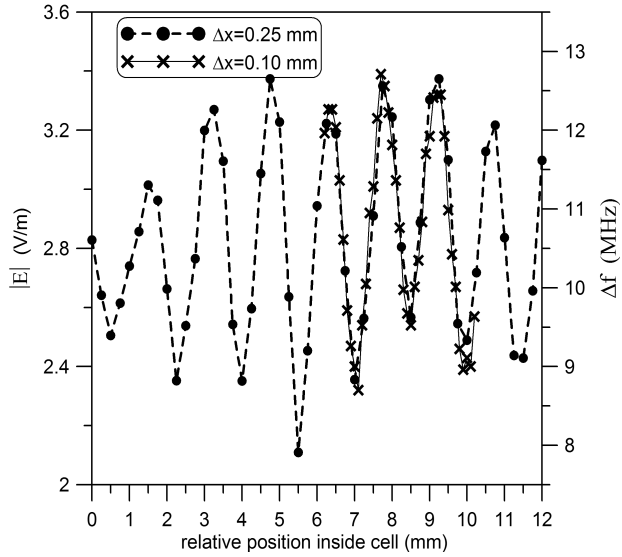}}
\caption{Measured EIT splitting, $\Delta f$ (right axis) and corresponding electric field amplitude $|E|$ (left axis) as a function of position inside the cell at 104.77~GHz. }
\label{splitting104}
\end{figure}

The cell is translated away from the source in discrete steps.
At each step position we measure the splitting of the EIT signal.  We convert to an electric field using Eq.~\ref{mage2}, and the dipole matrix element for this transition, $\wp = 473.14ea_0$.  Here, $e$ is the electron charge and $a_0$ is the Bohr radius.
The dashed line in Fig.~\ref{splitting104} shows the measured splitting (right axis) and the E-field amplitude (left axis) as a function of position for a step size of 0.25~mm. The crosses show a second scan at higher resolution with a step size of 0.10~mm, which corresponds to the larger of the two laser beam widths.  The origin in Fig.~\ref{splitting104} corresponds to a distance of approximately 8.4~mm between the laser beams and the inside edge of the cell that is furthest from the source.

These two sets of measurements lie on top of one another, showing the measurement is repeatable.  The results further demonstrate significant periodic field variation inside the cell. We see up to approximately $\pm 20~\%$ variation in the field amplitude over the 12~mm scan. The average period of the observed pattern is approximately half the wavelength of the RF ($\lambda_{RF} /2 = 1.43$~mm). The two sets of measurements yield field-mapping resolutions of $\approx \lambda/10$ and $\approx \lambda/30$, respectively.  We have thus shown that this method enables sub-wavelength mapping of 104.77~GHz radiation fields, and, importantly, that the achieved spatial resolution is comparable to the laser beam spot size.

\begin{figure}[!t]
\centering
\scalebox{.24}
{\includegraphics*{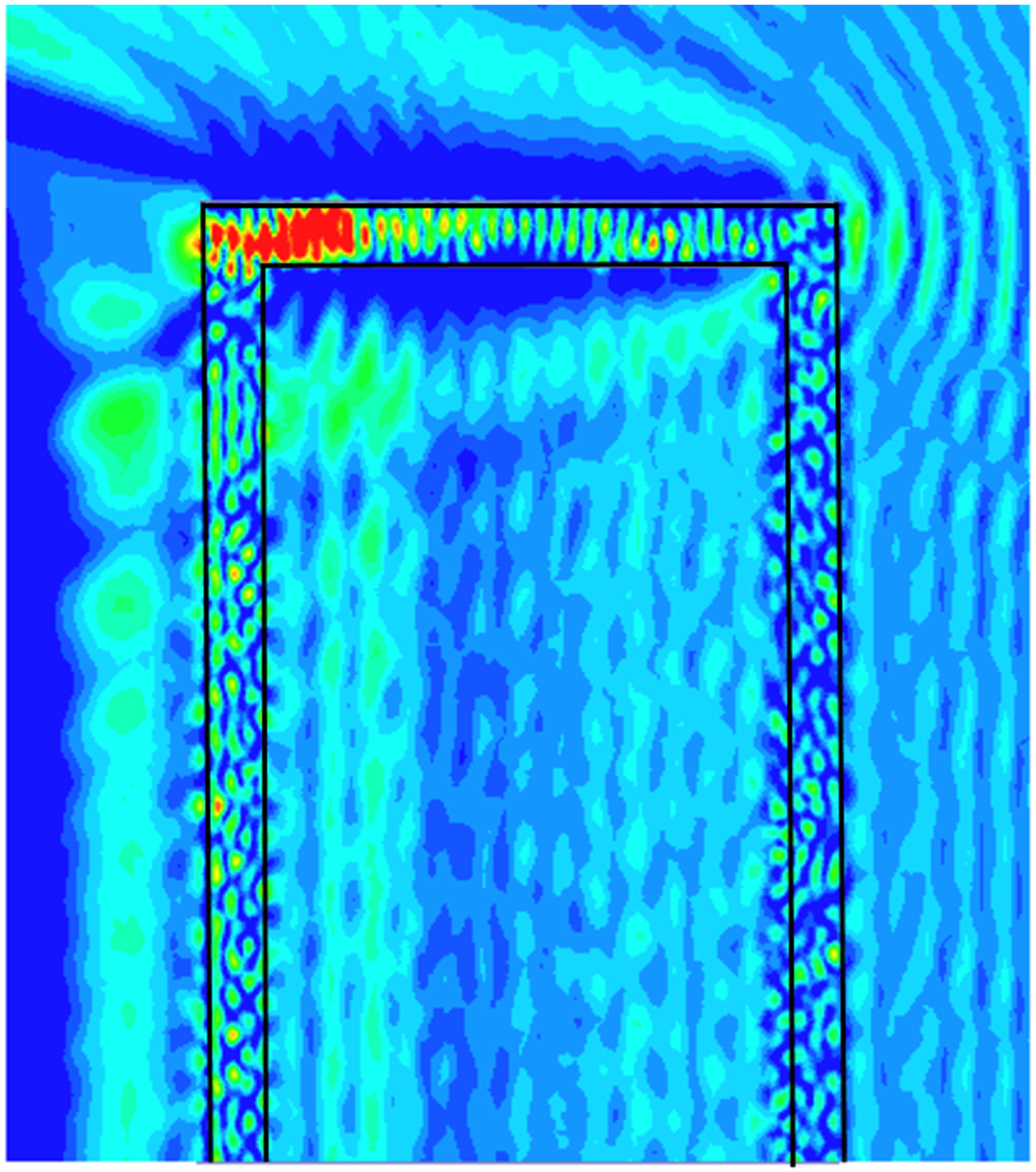}}
\scalebox{.24}
{\includegraphics*{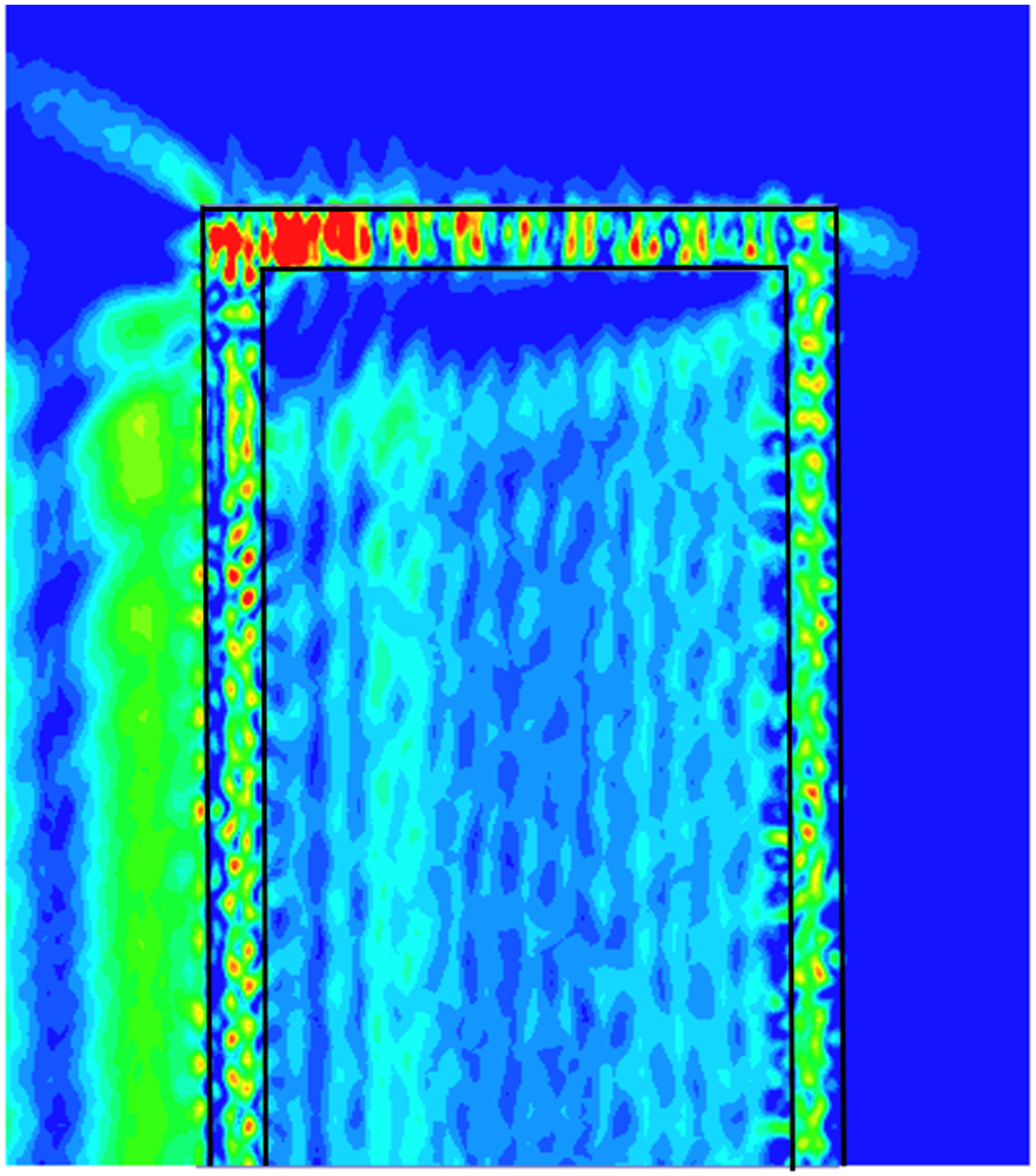}}\\
\footnotesize{(a) \hspace{30mm} (b)}
\caption{Simulation of electric-field amplitude $|E|$ for a plane wave incident onto the vapor cell from the right: (a) incident + scattered, and (b) scattered. The region shown in the figure is a horizontal planar cut through the center of the cell, with half of the cell shown. The E-field is on a linear colorscale ranging from 0.7 V/m (blue) to 5 V/m (red). }
\label{contour}
\end{figure}

\begin{figure}
\centering
\scalebox{.3}
{\includegraphics*{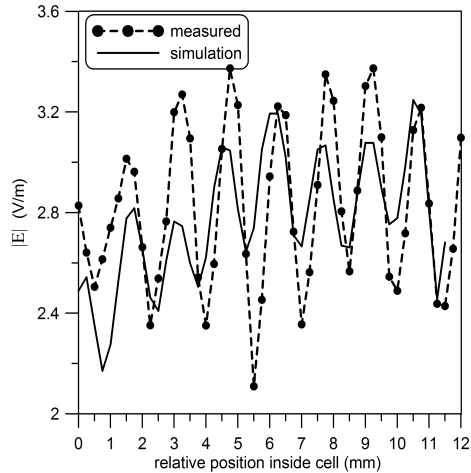}}
\caption{Comparison of experimental and simulated $|E|$ as a function of position inside the cell at 104.77~GHz.}
\label{fieldcut}
\end{figure}

Next, we compare the measured field distribution inside the cell to the results of a three-dimensional numerical full wave simulation performed using a commercial finite element code. It is challenging to perform such a numerical simulation of the actual experimental setup at high frequencies because of computer memory requirements. This is partially due to the small RF wavelength at 104.77~GHz, and the relatively large cell size (several RF wavelengths at 104.77~GHz). To overcome these issues we did the following. Instead of modeling the actual open-ended waveguide placed 0.14~m from the cell, we performed a numerical simulation for a plane-wave impinging on the cell with a field strength of 2.8~V/m.  In order to determine this field-strength value for the plane-wave, the open-ended waveguide source antenna was modeled independently and the numerical simulated far-field radiation pattern and field strength was determined. Using the measured working distance and power mentioned above, this yields a field amplitude of 2.8~V/m at the location of the laser beams in the experiment. The cell in the simulation has dimensions as mentioned above for the experiment, with glass having $\epsilon_r=5.5$ at RF frequencies. We assumed $\epsilon_r=1$ for the region inside the cell. Figure~\ref{contour} shows the numerical results for the field inside the cell with the incident plane wave source. This contour plot shows the expected field distribution, in which the field variation is primarily along the propagation direction of the incident RF wave. This indicates that our measurement method gives good spatial resolution in the only dimension along which there is significant field strength variation. To quantitatively compare simulation with measurement, we average the numerical results along 92~$\%$ of the length of the cell. The comparison is shown in Fig.~\ref{fieldcut}, which shows that the simulation and measurements give similar spatial variation.  Both show field distributions with a period roughly equal to half the RF wavelength.  We see good qualitative agreement between the numerical results and the data.

\begin{figure}[!t]
\centering
\scalebox{.3}
{\includegraphics*{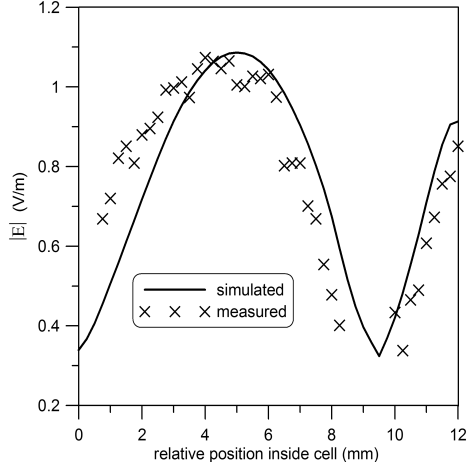}}
\caption{Experimental and simulated $|E|$ as a function of position inside the cell at 17.04~GHz.}
\label{field17}
\end{figure}

To study field mapping in the cm-wave regime we repeat the experiment with a 17.04~GHz source.  Here, the blue laser is tuned to $\approx 480.13$~nm to couple states $5P_{3/2}$ and $50D_{5/2}$.  The RF couples states $50D_{5/2}$ and $51P_{3/2}$.  We use a horn antenna at a distance of 0.880~m from the laser beams. Figure~\ref{field17} shows the measured E-field as a function of position inside the cell, where we have used $\wp = 1574.83ea_0$. For this measurement the origin of the position axis corresponds to the laser beams being approximately 10~mm from the inner surface of the cell that is furthest from the RF source.
The variation of the measured field is approximately $\pm 50$~$\%$ of its average, and the observed separation between the maximum and minimum is $\approx \lambda_{RF}/4$.

We perform numerical simulations for this case as well, using a plane wave source with a field amplitude at the vapor cell determined from a far-field calculation. Based on source power, cable losses, and known antenna characteristics, the field amplitude at the location of the laser beams is 0.76~V/m. In Fig.~\ref{field17}, we compare the data with results obtained from the numerical simulation. Here, agreement is good.

\section{Discussion and Conclusion}

In this paper we have demonstrated a technique based on Rydberg atoms, EIT, and Autler-Townes splitting which can perform sub-wavelength imaging and field mapping of RF radiation.  We have validated the approach by comparing the measured field inside a hollow glass cylinder to results obtained from a full-wave numerical simulation.

While the spatial resolution of our measurements is determined by the $\approx 100 \mu$m beam widths of our lasers, the spatial resolution of the method is in principle limited by the optical diffraction limit.  This is a significant improvement over the measurement resolution achievable by conventional probes.
There are many possible applications of this technique. For example, the sensing volume could be scanned over a printed-circuit-board (PCB) or a metasurface in order to map their fields, as well as other applications where E-field measurements on a small spatial resolution are desired. We aim to demonstrate these applications in future work.

\end{document}